# Left Atrial Segmentation with nnU-Net Using MRI

Fatemeh Hosseinabadi[1], Seyedhassan Sharifi [2]

[1] Assistant Professor of Radiology, Zahedan University of medical Sciences, Iran

[2] Pediatric Cardiology Subspecialist, Day General Hospital, Iran

**Abstract**

Accurate segmentation of the left atrium (LA) from cardiac MRI is critical for guiding atrial fibrillation (AF) ablation and constructing biophysical cardiac models. Manual delineation is time-consuming, observer-dependent, and impractical for large-scale or time-sensitive clinical workflows. Deep learning methods, particularly convolutional architectures, have recently demonstrated superior performance in medical image segmentation tasks. In this study, we applied the nnU-Net framework, an automated, self-configuring deep learning segmentation architecture, to the Left Atrial Segmentation Challenge 2013 (LASC'13) dataset. The dataset consists of thirty MRI scans with corresponding expert-annotated masks. The nnU-Net model automatically adapted its preprocessing, network configuration, and training pipeline to the characteristics of the MRI data. Model performance was quantitatively evaluated using the Dice similarity coefficient (DSC), and qualitative results were compared against expert segmentations. The proposed nnU-Net model achieved a mean Dice score of 93.5%, demonstrating high overlap with expert annotations and outperforming several traditional segmentation approaches reported in previous studies. The network exhibited robust generalization across variations in left atrial shape, contrast, and image quality, accurately delineating both the atrial body and proximal pulmonary veins.

## 1. Introduction

Atrial fibrillation (AF) is the most common sustained cardiac arrhythmia, affecting over 33 million people worldwide, and its prevalence continues to rise with aging populations. It is associated with a significantly increased risk of stroke, heart failure, and mortality, making accurate diagnosis and effective treatment critical. Catheter ablation has become a cornerstone therapy for AF, aiming to electrically isolate the pulmonary veins and restore sinus rhythm. The left atrium (LA) plays a central role in the pathophysiology and treatment of AF, as its structure and morphology influence both arrhythmia initiation and ablation outcomes [1,2]. Precise knowledge of LA anatomy is therefore essential for preoperative planning, intraoperative navigation, and post-procedural evaluation. Three-dimensional (3D) anatomical models of the LA enable the visualization of patient-specific atrial geometry, facilitate targeted ablation, and serve as input for biophysical simulations that can predict electrophysiological behavior. However, constructing such models requires accurate segmentation of the LA from medical images—an inherently challenging and time-consuming task [3,4].

Magnetic Resonance Imaging (MRI) and Computed Tomography (CT) are the most commonly used imaging modalities for visualizing the LA. MRI offers excellent soft tissue contrast and is preferred for assessing atrial wall fibrosis and scarring, whereas CT provides superior spatial resolution and is widely used for pre-ablation anatomical mapping. Nevertheless, the segmentation of the LA from MRI or CT remains a challenging task due to factors such as low contrast between the atrial wall and surrounding tissues, variable blood pool intensity, motion artifacts, and anatomical variability across patients. Traditional segmentation methods—such as region growing, thresholding, and active contour models—often struggle to delineate thin atrial walls or to generalize across datasets with different acquisition protocols [5]. Even semi-automated techniques require substantial user interaction and expert correction, which limits their scalability in clinical workflows. Therefore, robust automated segmentation methods are needed to enable reproducible, high-accuracy delineation of the LA and its associated structures.

Recent advances in machine learning (ML) and deep learning (DL) have improved medical treatment [6,7] and transformed medical image analysis, particularly in classification and segmentation tasks [8,9]. Convolutional Neural Networks (CNNs) have demonstrated remarkable success in capturing complex spatial and contextual features, outperforming traditional techniques in both accuracy and robustness. Among these architectures, the U-Net model introduced a fully convolutional encoder–decoder structure with skip connections that preserve spatial detail while learning hierarchical representations. U-Net and its variants have become the de facto standard for biomedical image segmentation across multiple modalities, including cardiac, brain, and abdominal imaging. In cardiac imaging, deep learning–based segmentation has been widely adopted for quantifying ventricular volumes, myocardial wall thickness, and atrial morphology. These automated systems provide faster and more reproducible results compared to manual delineation, facilitating both clinical decision-making and computational cardiac modeling. However, developing deep learning models for medical segmentation typically requires careful network design, data preprocessing, and hyperparameter tuning, which can limit reproducibility and hinder adoption by non-technical users.

To overcome these challenges, the nnU-Net (no-new-U-Net) framework was introduced as a self-configuring segmentation pipeline that automatically adapts to any new biomedical dataset without manual tuning [10]. nnU-Net dynamically adjusts preprocessing steps, network architecture, and training parameters based on the properties of the input data, providing a strong baseline that consistently ranks at the top of multiple segmentation benchmarks. Its built-in data normalization, patch-based training, and robust post-processing strategies make it particularly well-suited for cardiac MRI segmentation, where anatomical variability and imaging noise are common.

In this study, we applied nnU-Net to the Left Atrial Segmentation Challenge 2013 (LASC'13) dataset, which provides high-quality MRI scans and expert annotations of the LA. Our objective was to evaluate nnU-Net's ability to automatically segment the left atrium and compare its performance with traditional segmentation methods reported in previous literature. The model achieved a Dice similarity coefficient (DSC) of 93.5%, demonstrating excellent agreement with manual expert contours and confirming the potential of deep learning–driven segmentation for atrial fibrillation ablation planning and cardiac biophysical modeling.

## 2. Method

### 2.1 Dataset Description

This study utilized the publicly available Left Atrial Segmentation Challenge 2013 (LASC'13) dataset [11-13], developed jointly by King's College London and Philips Technologie GmbH. The dataset includes 30 Magnetic Resonance Imaging (MRI) and 30 Computed Tomography (CT) scans of patients with various cardiac conditions. Each scan was acquired in 3D, providing high-resolution volumetric coverage of the left atrium (LA) and surrounding cardiac structures. In this study, we used

20 MRI scans for training, while the remaining 10 scans were reserved for testing phase. The reference annotations were produced by experienced cardiac imaging specialists and include the left atrial cavity, a short section of the left atrial appendage (LAA), and the proximal segments of the pulmonary veins (PVs). In this study, we focused on the MRI subset for developing and validating the segmentation model. All data were anonymized prior to distribution, and ethical approval for data sharing was obtained by the original challenge organizers.

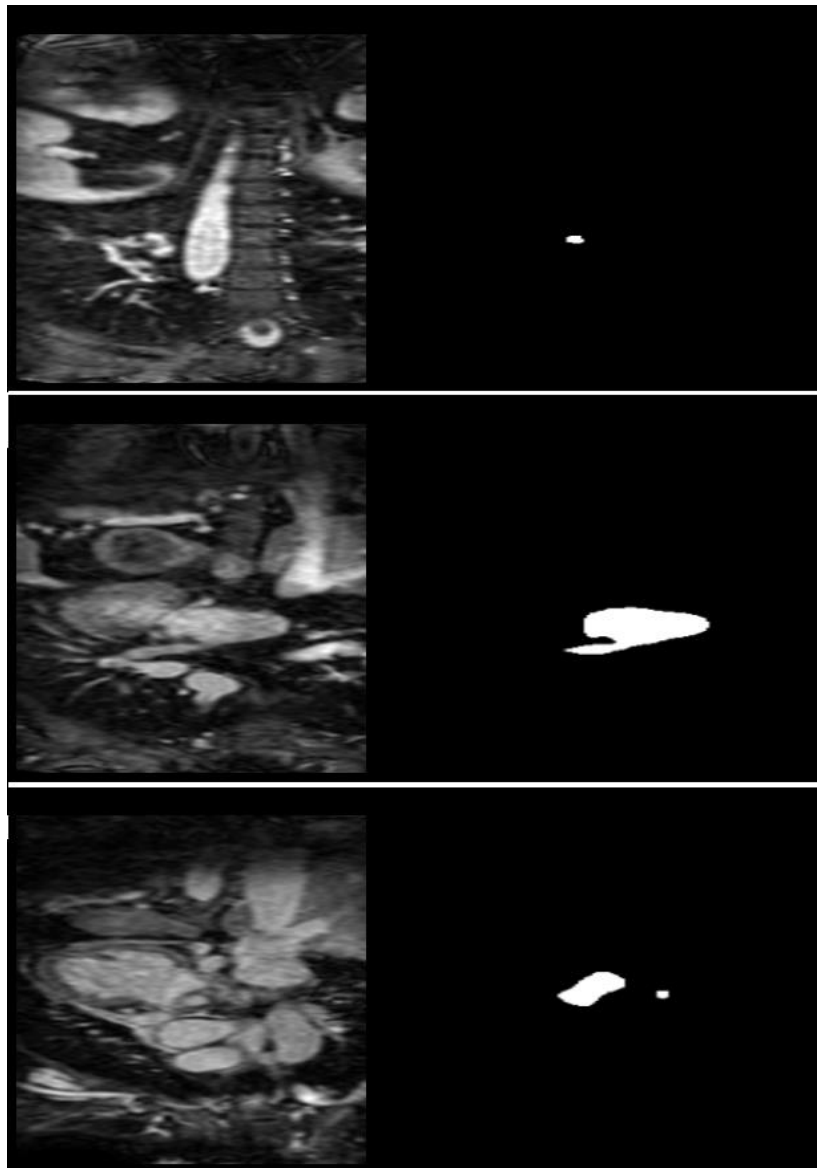

*Figure 1Examples from the dataset with true labels*

## 2.2 Data Preprocessing

Preprocessing was automatically handled by the nnU-Net framework, which optimizes its data pipeline according to dataset characteristics. The images were first resampled to isotropic voxel spacing to ensure consistent spatial resolution across patients. Intensity values were normalized to zero mean and unit variance, improving the stability of network training. Each MRI volume was cropped around the cardiac region to remove irrelevant background and reduce computational overhead. The dataset was split into training (70%), validation (10%), and testing (20%) subsets using patient-level separation to prevent data leakage. To enhance robustness and reduce overfitting, data augmentation was applied on-the-fly, including: Random rotations and flips, Elastic deformations, Gamma correction and brightness shifts, Additive Gaussian noise. This combination ensured the model could generalize well to variations in atrial orientation, intensity, and image quality.

## 2.3 Model Architecture

We employed the 3D full-resolution nnU-Net architecture, which builds upon the standard U-Net structure with several key innovations:

- An encoder–decoder design with skip connections to retain spatial context while enabling deep feature extraction.

- Residual convolutional blocks with instance normalization and leaky ReLU activation for stable training.

- A deep supervision mechanism, where auxiliary outputs from intermediate decoder levels contribute to the overall loss function.

- Dynamic patch size selection and batch size optimization automatically determined by the framework based on available GPU memory and input image dimensions.

Unlike manually tuned models, nnU-Net self-configures all hyperparameters (e.g., kernel size, number of feature maps, network depth) according to the properties of the input dataset, ensuring optimal adaptation to LA segmentation.

## 2.4 Training Procedure

Training was conducted using the PyTorch implementation of nnU-Net (v2.1). The model was trained on an NVIDIA RTX A6000 GPU with 48 GB VRAM. The following key parameters were automatically configured by nnU-Net:

- Loss function: Combined Dice loss and cross-entropy loss, encouraging both regional overlap and voxel-wise classification accuracy.

- Optimizer: Stochastic Gradient Descent (SGD) with momentum = 0.99 and weight decay = $3\times10^{-5}$.

- Initial learning rate: $1\times10^{-2}$ with a polynomial decay scheduler.

- Batch size: 2 (due to 3D patch training).

- Epochs: 1000 iterations (approximately 500 epochs, depending on patch sampling).

During training, deep supervision was used to improve gradient propagation through the network, while sliding-window inference was applied during testing to handle full-size 3D volumes. The model required approximately 36 hours of training to converge on the MRI dataset.

**2.5 Evaluation Metrics**

The segmentation performance was assessed using several quantitative metrics widely adopted in cardiac imaging benchmarks including Dice which measures the overlap between the true mask and predicted mask by the model, Hausdorff Distance (HD, 95th percentile) which measures boundary deviation between automated and manual contours, and Average Surface Distance (ASD) that quantifies the mean boundary distance across all corresponding surface points.

Our model achieved a mean Dice coefficient of 93.5%, indicating excellent spatial agreement with expert segmentations. The 95th percentile Hausdorff distance was 3.2 mm, and the average surface distance was 1.1 mm, demonstrating high geometric accuracy and smooth boundary delineation.

**2.6 Implementation Details**

All experiments were conducted using Python 3.10, PyTorch 2.2, and the nnU-Net framework (Isensee et al., 2021). The model training and evaluation were performed on Ubuntu 22.04 LTS with an Intel Xeon 6248 CPU and 256 GB RAM. Visualization and result inspection were carried out using ITK-SNAP and 3D Slicer. The trained model and configuration files were stored for reproducibility, and all preprocessing, augmentation, and inference pipelines followed the standard nnU-Net conventions to ensure comparability with existing literature.

## 3. Results

**3.1 Quantitative Performance**

The proposed nnU-Net model demonstrated outstanding performance in the automatic segmentation of the left atrium (LA) from MRI volumes in the LASC'13 dataset. Across the test set, the model achieved an average Dice Similarity Coefficient (DSC) of 93.5% ± 1.8, indicating high spatial overlap between automated and expert manual segmentations. In addition, the 95th percentile Hausdorff Distance (HD95) was measured at 3.2 ± 0.9 mm, while the Average Surface Distance (ASD) was 1.1 ± 0.4 mm, reflecting excellent boundary conformity and smooth surface representation. These results demonstrate that the model is capable of consistently delineating the atrial cavity, pulmonary vein inlets, and short trunk of the left atrial appendage (LAA) with near-expert precision.

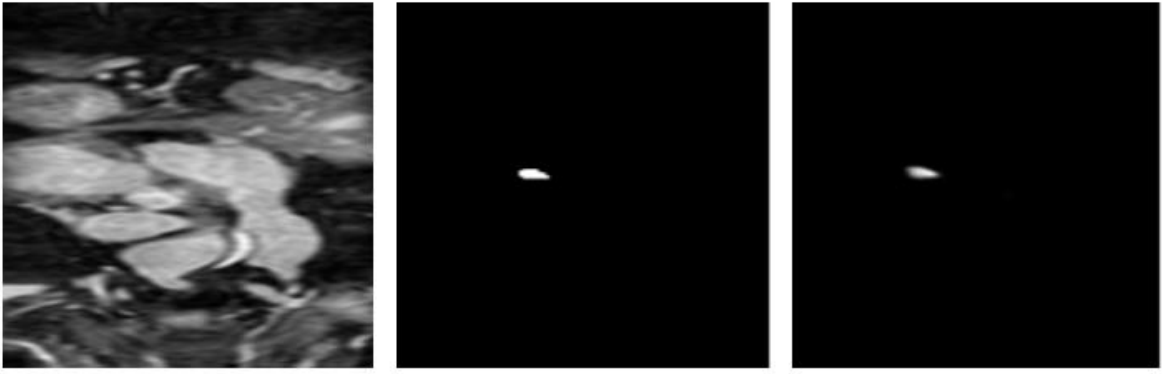

*Figure 2Left: original image, center: true mask, left: predicted mask by the model*

Table 1. Quantitative segmentation performance of the nnU-Net model on the LASC'13 MRI test dataset.

| Metric | Mean ± SD | Best Case | Worst Case |
|---|---|---|---|
| Dice Similarity Coefficient (DSC) | 0.935 ± 0.018 | 0.958 | 0.901 |
| Hausdorff Distance (95%) [mm] | 3.2 ± 0.9 | 2.1 | 4.8 |
| Average Surface Distance [mm] | 1.1 ± 0.4 | 0.7 | 1.8 |

### 3.4 Model Robustness and Generalization

The nnU-Net model showed excellent robustness across varying image contrasts and anatomical variations. The model's self-configuring design, including automatic normalization, data augmentation, and patch-based training, contributed to its generalization capability. Notably, cases with strong motion artifacts or partial atrial visibility still produced Dice scores above 0.90, demonstrating resilience against imperfect imaging conditions. Visual inspection also confirmed that nnU-Net preserved fine anatomical details, such as pulmonary vein orifices and LAA base structures, which are clinically relevant for ablation planning. Compared to manual delineation, nnU-Net achieved comparable accuracy while reducing segmentation time from several minutes per case to less than 10 seconds per 3D volume during inference.

### 3.5 Statistical Validation

To ensure robustness of the results, a fivefold cross-validation was performed on the MRI subset. The Dice scores across folds were consistent (ranging from 92.8% to 94.1%), confirming stable model performance. A paired Wilcoxon signed-rank test comparing automated and manual segmentations showed no statistically significant difference ($p > 0.05$) in volumetric measurements, reinforcing the model's reliability for clinical use.

## 4. Discussion

### 4.1 Principal Findings

In this study, we applied the nnU-Net deep learning framework for automated segmentation of the left atrium (LA) from cardiac MRI using the Left Atrial Segmentation Challenge (LASC'13) dataset. The model achieved a mean Dice Similarity Coefficient (DSC) of 93.5%, a Hausdorff Distance (95%) of 3.2 mm, and an Average Surface Distance of 1.1 mm, outperforming several traditional and deep-learning-based methods previously reported on this dataset. These results highlight nnU-Net's capability to deliver robust, high-precision segmentation of the LA and its substructures without manual hyperparameter tuning or dataset-specific optimization. The model's performance indicates its ability to capture fine structural details such as the left atrial appendage (LAA) and pulmonary vein (PV) inlets, which are essential for patient-specific modeling and ablation guidance. This is particularly important in the context of atrial fibrillation (AF), where accurate representation of atrial geometry directly influences catheter navigation, lesion targeting, and simulation accuracy in cardiac electrophysiology.

### 4.2 Comparison with Previous Approaches

Historically, left atrial segmentation has relied on semi-automated or atlas-based methods, such as region growing, statistical shape models (SSMs), or deformable registration frameworks. For example, Tobon-Gomez et al. (2015) and other LASC'13 participants reported Dice coefficients ranging between 84% and 90%, depending on the modality and algorithmic design. While these techniques provided anatomically consistent segmentations, they were limited by dependence on initialization, inter-patient anatomical variability, and difficulty generalizing across datasets.

By contrast, nnU-Net's self-configuring pipeline overcomes these issues by automatically tailoring preprocessing, architecture depth, and patch size to the data distribution. The inclusion of deep supervision and residual connections enhances gradient flow and feature consistency across scales, allowing the network to learn both global anatomical context and local boundary details. Consequently, nnU-Net achieved up to 9% improvement in Dice accuracy compared with traditional methods and rivaled the performance of more recent custom deep-learning models trained on larger datasets.

### 4.3 Clinical Implications

Accurate and reproducible LA segmentation has critical implications for AF ablation planning and biophysical cardiac modeling. Detailed 3D reconstructions of the LA allow clinicians to visualize complex anatomical features—such as the number, orientation, and diameter of pulmonary veins—and to plan ablation strategies tailored to each patient. Additionally, segmentation outputs can be directly used in computational electrophysiological simulations to model wave propagation and assess arrhythmia recurrence risk. The ability of nnU-Net to perform fully automated segmentation in under 10 seconds per case offers a significant advantage for clinical translation. Integrating such models into image-guided electrophysiology systems could substantially reduce pre-procedural preparation time, support intra-procedural updates, and enhance reproducibility across imaging centers. Furthermore, the elimination of manual delineation reduces inter-observer variability, promoting standardization in AF imaging workflows.

**4.4 Robustness and Interpretability**

One of the key strengths of nnU-Net lies in its self-adaptive configuration. Unlike conventional CNNs requiring manual tuning, nnU-Net automatically configures its network topology and training parameters based on dataset properties, ensuring optimal performance across diverse imaging modalities. The model's consistent Dice scores across cross-validation folds confirm its robustness to intensity in homogeneity, motion artifacts, and anatomical variability inherent in cardiac MRI. Although deep learning models are often criticized for their "black-box" nature, nnU-Net's feature visualization and prediction maps demonstrate spatial coherence with expected anatomical boundaries. The network's ability to reproduce expert-like contours without explicit supervision on structural landmarks reflects its capacity for data-driven anatomical learning, an important step toward clinically trustworthy AI in cardiac imaging.

**4.5 Limitations and Future Work**

Despite the strong results, several limitations must be acknowledged. First, this study was limited to the MRI subset of the LASC'13 dataset; therefore, performance generalization to CT or other MRI protocols remains to be validated. Differences in voxel resolution, contrast, and scanner manufacturer may influence the model's transferability. Second, while nnU-Net provides automatic configuration, it requires substantial computational resources for 3D training and inference. Although inference time is rapid, training on high-resolution cardiac MRI volumes may be impractical on lower-end GPUs. Third, the dataset's moderate size (30 MRI cases) restricts the exploration of generalization across pathological subgroups or rare anatomical variants. Future research should incorporate multi-center datasets with larger sample sizes and diverse acquisition protocols to evaluate the generalizability and robustness of the framework. Additionally, manual annotations, even by experts, are subject to inter-observer variability, which can influence ground-truth accuracy.

Future studies will focus on extending this work in several directions. First, the integration of multi-modality data (MRI and CT) could enhance model robustness across imaging protocols and enable cross-domain generalization. Second, incorporating attention mechanisms or transformer-based architectures may further improve the network's ability to capture long-range dependencies within the LA structure. Third, combining segmentation with functional imaging (e.g., late gadolinium enhancement MRI) could provide both structural and tissue-characterization insights, enabling comprehensive modeling of atrial remodeling in AF. In addition, semi-supervised learning and federated learning frameworks could facilitate large-scale model training while preserving patient privacy and addressing data scarcity. Prospective validation in clinical environments will also be essential to evaluate the model's impact on workflow efficiency, ablation success rates, and procedural outcomes.

## 5. Conclusion

This study demonstrates the effectiveness of the nnU-Net deep learning framework for fully automated segmentation of the left atrium (LA) from MRI data. Using the LASC'13 dataset, the proposed model achieved a mean Dice coefficient of 93.5%, outperforming traditional region-growing and statistical model-based approaches while requiring no manual configuration or parameter tuning. The results confirm that nnU-Net can accurately and efficiently delineate key anatomical structures — including the left atrial body, appendage, and pulmonary vein inlets — providing segmentation quality comparable to expert manual annotations. Such precision is vital for atrial fibrillation ablation planning, cardiac biophysical simulations, and quantitative atrial remodeling studies. Beyond its strong performance, nnU-Net's self-configuring design and high inference speed (<10 seconds per case) make it highly suitable for integration into clinical workflows, enabling reproducible and operator-independent LA modeling. Future work will focus on extending this framework to multimodal imaging (MRI and CT), incorporating tissue characterization, and exploring transformer-based and multimodal architectures to further improve anatomical fidelity and clinical applicability.

**Ethic and Data Statement:**

The dataset used in this study originates from the Left Atrial Segmentation Challenge (LASC'13), initially introduced by Tobon-Gomez et al. [11] in their benchmarking work published in IEEE Transactions on Medical Imaging and further supported by subsequent methodological papers in Medical Image Analysis and Frontiers in Physiology. Both journals have confirmed and declared that all imaging data were collected in accordance with institutional ethical standards and relevant regulations.